\documentstyle[12pt]{article}
\begin{document}

\title
{\Large \bf Synchronization of Chaotic Maps by Symmetric Common Noise}

\author{ C.-H. Lai$^{1,2}$ and Changsong Zhou$^1$ \\
        $^1$Department of Computational Science\\
        and $^2$Department of Physics\\
        National University of Singapore,
        Singapore 119260}

\date{}
\maketitle

\begin{center}
\begin{minipage}{14cm}

\centerline{\bf Abstract}

Synchronization of identical chaotic systems subjected to common noise
has been the subject of recent research. Studies on several chaotic
systems have  shown that, the synchronization is actually induced by
the non-zero mean of the noise,  and  symmetric noise with zero-mean
cannot lead to synchronization.  Here  it is  presented that
synchronization can be achieved by {\sl zero-mean} noise in some
chaotic maps with large convergence regions.

\bigskip
\baselineskip 24pt

PACS number(s): 05.45.+b

\end{minipage}
\end{center}

\bigskip 

Running title: Synchronization by Noise

\newpage

The effect of common noise on the synchronization of identical chaotic
systems has attracted much attention since the work by Maritan and
Banavar[1], which claimed that two identical chaotic systems
subjected to the same strong enough noise can be synchronized, with the
logistic map and the Lorenz system as examples. Some authors have
reconsidered their conclusion.  Pikovsky[2] pointed  out that the
largest Lyapunov exponent of the noisy logistic map is positive which
is in contradiction to the criterion of negative largest Lyapunov
exponent for synchronization[3].  It was also pointed out[2] that the
synchronization observed in [1] is an outcome of finite precision in
numerical simulations, which was further confirmed by a detailed
study by Longa {\sl et al}[4].

Several other authors, on the other hand, reconsidered the problem by
examining the properties of the noise added to the systems. For the
case of noisy logistic map
\begin{equation}
x_{n+1}=4x_n(1-x_n)+\xi,
\end{equation}  
where the random number $\xi$ is chosen from the interval $[-W, W]$
with the constraint $x_{n+1}\in (0,1)$; otherwise, a new random number
is chosen.  Such a state-dependent noise is no longer symmetric[5], but
has a negative mean[6], and it is this nonzero mean that plays an
important role in the coalescence of trajectories.  
A zero-mean noise, although is still state-dependent,
cannot lead to synchronization[6].

For the case of the Lorenz system, synchronization  was observed by Maritan
and Banavar for uniform noise in $[0, W]$, but not for symmetric
noise. In [5], it was shown that the largest Lyapunov exponent of the
noisy Lorenz system is the same as that of the system driven constantly
by the mean value of the noise, indicating that the bias of the noise
plays the central role in synchronization. It was pointed out that the
origin of non-chaotic behavior is that the Lorenz system driven by
large enough constant perturbations is actually stable at the fixed
points[5,7].  Very recently, Sanchez {\sl et al}[8] analyzed the
synchronization of chaotic systems by noise in an experiment with the
Chua circuit, again drawing the conclusion that synchronization may
be achieved only by biased noise, and not symmetric noise.

So it seems that symmetric noise cannot convert a chaotic system into a
nonchaotic one, so that synchronization will occur for systems in
common noise.  In this letter, we are going to present an example that
synchronization can actually be achieved by symmetric, zero-mean common
noise.  In order to avoid the effect of the boundary of a system,
such as  the logistic map, on the realization of noise, we choose a
system that can be driven by noise of any level. The chaotic
systems driven by noise are  written as
\begin{eqnarray}
x_{n+1}&=& f(x_n)+\xi,\\
y_{n+1}&=& f(y_n)+\xi,
\end{eqnarray}  
with
\begin{equation}
f(x)=\tanh(A_1x)-B\tanh(A_2x),
\end{equation} 
where $A_1, A_2$ and $B$ are parameters of the map.  Chaos can  easily
occur in such a nonlinear system. With $A_1=20$ and $A_2=2$, the
Lyapunov exponent of the noise-free system
\begin{equation}
\lambda=\lim _{N\to \infty} \frac{1}{N}\sum\limits_{n=1}^N
\ln|f^{\prime}(x_n)|, 
\end{equation}
is calculated as a function of $B$, as shown in Fig. 1(a). Chaos occurs
in large regions of $B$ with $\lambda>0$.  In the following
simulations,  $B$ is fixed at 1.5, and the corresponding chaotic map is
shown in Fig. 1(b).

For the noisy chaotic system, the Lyapunov exponent can be estimated by
exactly the same formula as Eq. (5), but $\{x_n\}$ is now a noise
trajectory.  For synchronization to occur, $\lambda$ should be
negative. $\lambda<0$ is possible for an appropriate level of noise so
that the state of the system has higher probability to reach
$|f^{\prime}(x_n)|<1$. Such a  region
\begin{equation}
C=\{x|\;\;\; |f^{\prime}(x)|<1\}
\end{equation}
is called the convergence region of the system, and is employed to
realize entrainment and migration control of chaotic systems[9].
 
Our first simulation is to evaluated the Lyapunov exponent of the noisy
system.  The noise is simulated with uniform random number having a
zero-mean and a variance $\sigma^2$.  With 100 random initial
conditions for the same realization of noise, the averaged Lyapunov
exponent is estimated as a function of $\sigma^2$, as shown in Fig. 2.
Synchronization with $\lambda<0$ occurs for $\sigma^2>0.24$. So,
in contrast to the examples in [1-8], the sensitivity of this chaotic
map can be suppressed by {\sl zero-mean} noise, so that  systems
starting from different initial conditions will finally collapse into
the same final orbit.

The synchronization can be understood from the view point of the
convergence region of the map.  We perform two calculations:  one
is the distribution of the state  of the system; and the other the
distribution of the finite-time Lyapunov exponents defined as[5]
\begin{equation}
\lambda^{(m)}=\frac{1}{m}\sum\limits_{n=1}^m \ln|f^{\prime}(x_n)|,
\end{equation}
which measures the average expansion  or contraction rate in $m$ steps.
The results for noise-free system($\sigma^2=0$), chaotic noisy system
($\sigma^2=0.1$) and non-chaotic noisy system ($\sigma^2=0.3$) are
shown in Fig. 3 (a) for the distribution of the state,  and Fig. 3(b)
for the distribution of $\lambda^{(10)}$.  In Fig. 3(a), the plot of
$|f^{\prime}|<1$ (solid line) is superimposed onto the normalized
distribution of $x_n$. It is seen that for the noise-free case, the
system spends most of its time out of the convergence region, and two
slightly different orbits will diverge after almost any 10 iterations,
because it is very rare that $\lambda^{(10)}<0$.  When the system is
subjected to  noise, it is driven to spend more time  in the
convergence region.  However, if the noise level is not high enough so
that temporal convergence is overcome by divergence in the system
dynamics, the system will remain chaotic, and synchronization will not
occur. When the noise level is higher then some threshold, the
sensitivity is suppressed, and synchronization is achieved.

Our next simulation is to examine the biased noise on synchronization.
A  biased noise can be denoted by its mean value and a symmetric noise
as
\begin{equation}
\xi_{asym}=a_v+\xi_{sym},
\end{equation}
where $\xi_{sym}$ is a  uniform random number with  mean value 0 and
variance $\sigma^2$.  For the Lorenz system studied in [1], it is shown
in [5,6] that, the largest Lyapunov of the noise system is the same
when the fluctuations $\xi_{asym}$ are replaced by the mean value $a_v$
of the noise, showing that it is the non-zero mean which plays the
central role in the synchronization of the system. So in our
simulations, the Lyapunov exponent is estimated as a function of $a_v$
for the noise system
\begin{equation}
x_{n+1}=f(x_n)+\xi_{asym},
\end{equation}
and the constantly driven system
\begin{equation}
x_{n+1}=f(x_n)+a_v.
\end{equation}  
The results for Eq. (9) with $\sigma^2=0.3$ and Eq. (10)  are shown in
Fig. 4.  The system of Eq. (10) is  still chaotic for most of  $a_v$ in
the region $a_v \in (0,0.5)$.  The noise system, however, is nonchaotic
for all the values of $a_v$.  In fact, the  constantly driven system of
Eq. (10) can be viewed  as a new chaotic system $f_1(x_n)=f(x_n)+a_v$,
and the system subjected to biased noise can be regarded as a  new
chaotic  system driven by symmetric noise.
\begin{equation}
x_{n+1}=f_1(x_n)+\xi_{sym},
\end{equation}
Again, synchronization is achieved by zero-mean noise. 

Similar behavior is observed for other parameters of the map.  We have
also studied other maps having similar large convergence regions, such
as the Gaussian map $f(x)=r\alpha x\exp(-2x^2+ax)$, where $\alpha=\sqrt{e}/2$.
For example, with $a=0$ and $r=7$, synchronization  can be shown to occur
for $\sigma^2>0.099$.

The counterintuitive effect of noise on chaotic map was reported in some
earlier studies[10-12] on the BZ map which is directly connected to the real
 chemical 
reaction, the Belousov-Zhabotinshy reaction, and has a similar structure
as the maps studied in this letter, i.e., having  steep and flat regions.
There, a small noise may change the chaotic orbit of the system into a 
state similar to a periodic orbit  with  noise[10], which leads to a negative 
Lyapunov exponent[10], a slower 
decay of correlations[11] and an improvement of  state predictability[12]. 
This noise-induce order was attributed to the steepness of the BZ map.
However, in our examples, the flat regions outside the original chaotic
attractors of the map play an important 
role in synchronization for large enough noise. Unlike the BZ map, the resulted
synchronized state is not similar to some periodic orbits.

Synchronization of chaotic systems has  potential applications in secure
communication[13]. However, information masked by low-dimensional chaos may
be attacked using some prediction-based methods[14-16]. Since
multiplicative noise may impose great difficulties in dynamical analysis,
 communication using synchronized noisy  maps may provide additional security. 
We are now investigating an appropriate realization of this idea.

In summary, we have shown that synchronization of identical chaotic
systems subjected to the same {\sl zero-mean} noise can be achieved for
large enough noise levels if the systems have  large convergence
regions.  Synchronization is achieved
not because of the constant bias of the noise, but because of the fact
that noise of a sufficiently high level drives the system deep into the
convergence region, where the difference between nearby trajectories
shrinks quickly. 
The general claim that synchronization of identical
chaotic system subjected to the same noise is induced by the nonzero
bias is not valid. 
This mechanism of synchronization is quite different from   
other synchronization approaches where synchronization is achieved when the  
invariant synchronization manifold is stable[17], thus providing a new
understanding of the behavior of driven nonlinear systems.  For future research, it is worthwhile to study the
synchronization of identical continuous chaotic systems subject to
common noise from the viewpoint of convergence regions, and application 
of the system in secure communication.

\bigskip
{\bf Acknowledgements:}
This work was supported in part by research grant RP960689 at the National
University of Singapore.  CZ is a NSTB Postdoctoral Research Fellow.

\newpage
\bigskip

{\bf References}
\begin{description}
\item [1.]  MARITAN A and  BANAVAR J. R, {\sl  Phys. Rev. Lett.} {\bf 72} (1994), 1451;
                                         {\sl Phys. Rev. Lett.} {\bf 73} (1994), 2932.
\item [2.]  PIKOVSKY A. S., {\sl Phys. Rev. Lett.} {\bf 73} (1994), 2931.

\item [3.]  YU L. , OTT E. and CHEN Q., {\sl Phys. Rev. Lett.} {\bf 65} (1990), 2935.
\item [4.]  LONGA L., CURADO E. M. F.  and OLIVEIRA F. A., 
             {\sl Phys. Rev. E} {\bf 54} (1996), R2201.
\item [5.]  HERZEL H.  and FREUND J., {\sl Phys. Rev. E} {\bf 52} (1995), 3238.
\item [6.]  MALESCIO G., {\sl Phys. Rev. E} {\bf 53} (1996), 6551.
\item [7.]  GADE P. M.  and BASU C., {\sl Phys. Lett. A} {\bf 217} (1996), 21.
\item [8.]  SANCHEZ E., MATIAS  M.A. and  PEREZ-MUNUZURI V., {\sl Phys. Rev. E} {\bf 56} (1997), 4068.
\item [9.]  JACKSON E. A., {\sl Phys. Lett. A} {\bf 151} (1990), 478.
\item [10.] MATSUMOTO K. and  TSUDA I., {\sl J. Stat. Phys.} {\bf 31} (1983), 87.
\item [11.] HERZEL H. P. and EBELING W., {\sl Phys. Lett. A} {\bf 111} (1985), 1.
\item [12.] HERZEL H. P. and POMPE B., {\sl Phys. Lett. A} {\bf 122} (1987), 121.
\item [13.] CUOMO K.M.and OPPENHEIM A.V., {\sl Phys. Rev. Lett.} {\bf 71} (1993),  65.
\item [14.] SHORT K. M., {\sl Int. J. Bifurcation Chaos} {\bf 4} (1994) 959.
\item [15.] PEREZ G. and  CERDERIA  H. A., {\sl Phys. Rev. Lett.}  {\bf 74} (1995), 1970 .
\item [16.] ZHOU C. S. and  CHEN T. L., {\sl Phys. Lett. A}  {\bf 234} (1997), 429.
\item [17.] PECORA L. M., CARROLL T. L., JOHNSON G.A., MAR D. J., and HEAGY
 J. F., {\sl Chaos} {\bf 7} (1997), 520. 
\end{description}

\bigskip
\newpage
{\large \bf Figure Captions}
\begin{description}
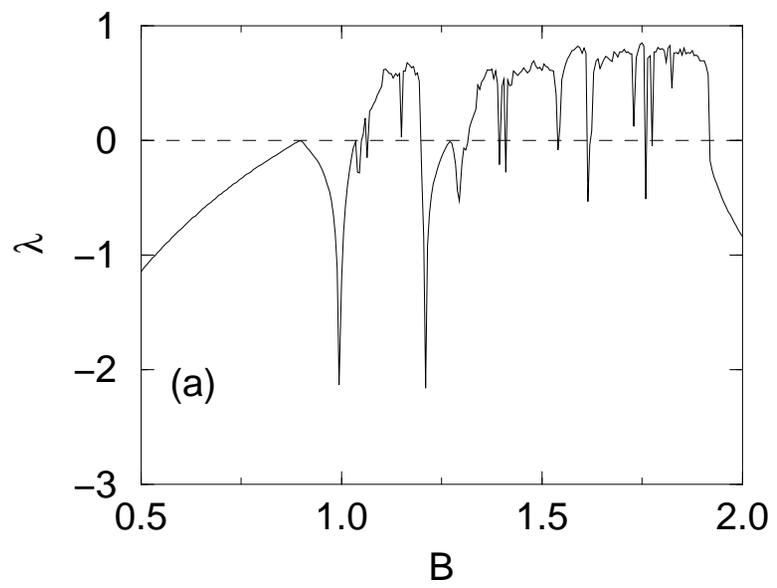
\item Fig. 1. (a) Lyapunov exponent of the map via the parameter $B$ at $A_1=20, A_2=2$.
              (b) The shape of the map for $A_1=20, A_2=2$ and $B=1.5$.

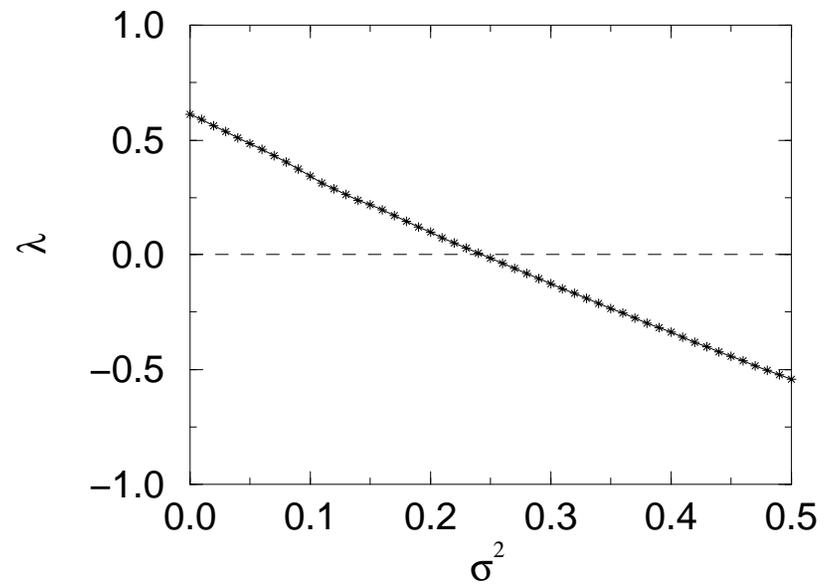
\item Fig. 2. Lyapunov of the noisy system as a function of noise level $\sigma^2$. 
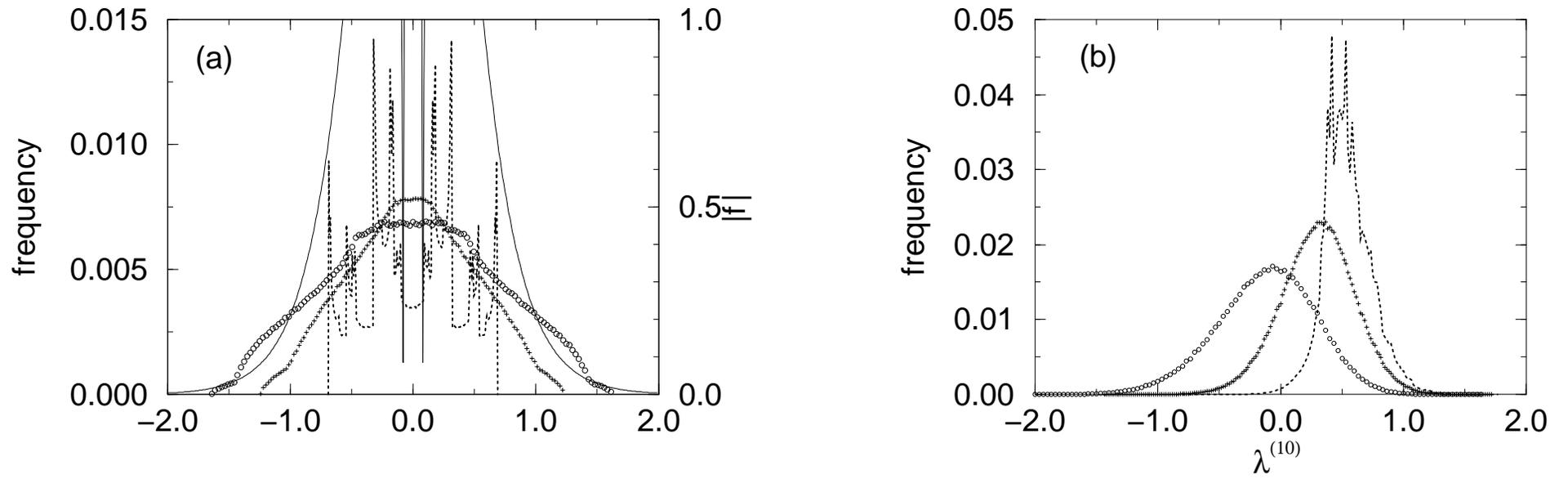
\item Fig. 3. Normalized histograms of the state $x_n$ (a),
              and finite-time Lyapunov exponent (b) for $\sigma^2=0$ (dots),
              $\sigma^2=0.1$  (pluses) and $\sigma^2=0.3$ (circles) from $5\times 10^6$ 
              iterations. The plot of
              $|f^{\prime}|<1$ (solid line) is superposited onto the figure (a) 
              to indicate the convergence regions.
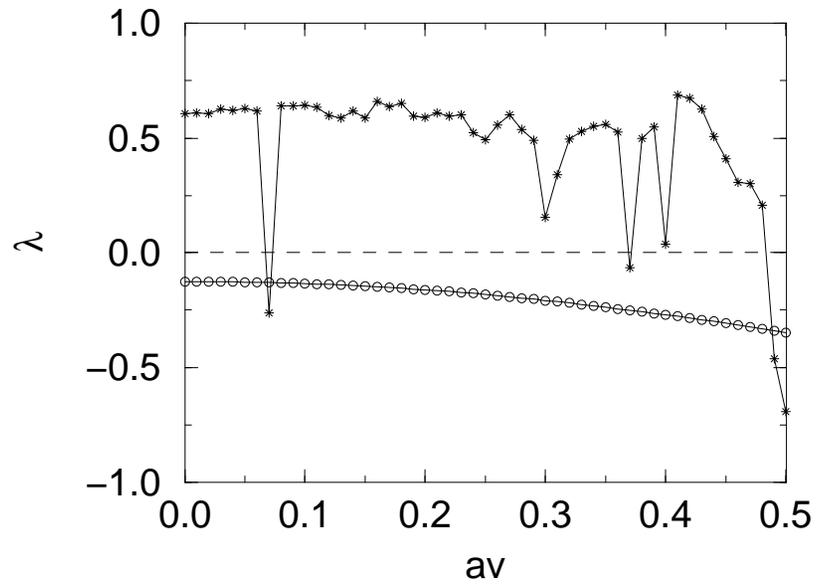
\item Fig. 4. Lyapunov exponents as a function of $a_v$. Circles, noise system 
              with $\sigma^2=0.3$; Stars, noise replaced by its mean value. 
\end{description}  
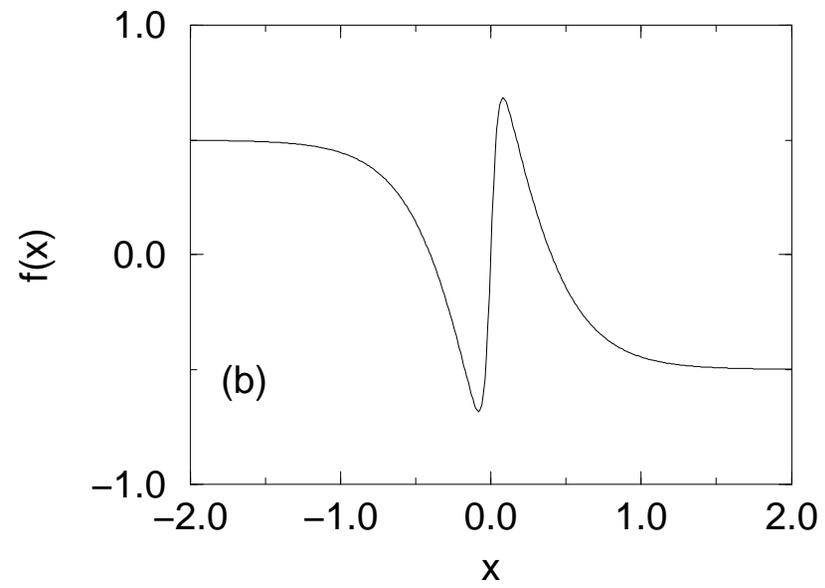
\end{document}